\documentclass[twocolumn,preprintnumbers,prl,amsmath,amssymb,superscriptaddress,showpacs]{revtex4}
\usepackage{epsfig}
\usepackage{graphicx}

\begin{document}

\title{Experimentally feasible quantum erasure-correcting code for continuous variables}

\author{J. Niset}
\affiliation{QuIC, Ecole Polytechnique, CP 165, Universit\'e Libre de
Bruxelles, 1050 Brussels, Belgium}

\author{U. L. Andersen}
\affiliation{Department of Physics, Technical University of Denmark, 
Building 309, 2800 Kgs. Lyngby, Denmark}
\affiliation{Institut f\"ur Optik, Information und Photonik, Max-Planck Forschungsgruppe, Universit\"at Erlangen-N\"urnberg,
Staudtstra{\ss}e 7/B2, 91058 Erlangen, Germany}

\author{N. J. Cerf}
\affiliation{QuIC, Ecole Polytechnique, CP 165, Universit\'e Libre de
Bruxelles, 1050 Brussels, Belgium}

\begin{abstract}
We devise a scheme that protects quantum coherent states of light from probabilistic losses, thus achieving the first continuous-variable quantum erasure-correcting code. If the occurrence of erasures can be probed, then the decoder enables, in principle, a perfect recovery of the original light states. Otherwise, if supplemented with postselection based on homodyne detection, this code can be turned into an efficient erasure-filtration scheme. The experimental feasibility of the proposed protocol is carefully addressed.

\end{abstract}

\maketitle

Transmitting, storing, or manipulating quantum information without errors is a main prerequisite to the realization of most quantum information processes. As errors are inherent to any realistic implementation, the future of quantum information systems strongly relies on the ability to detect and correct errors. While the theory of quantum error correction for two-level systems is well advanced (see, e.g., \cite{Shor}), very little is known if quantum information is encoded in continuous degrees of freedom such as the quadratures of a mode of light. The processing of quantum information based on such continuous variables (CV) is, however, very attractive as it offers the advantage of being relatively easy to implement. Many tasks such as the preparation of entangled states \cite{entangled}, quantum teleportation \cite{teleportation} or entanglement swapping \cite{swapping}, to name just a few, have been realized with optical parametric oscillators, beam splitters, and homodyne detection only. Since the pioneering works of Refs.~\cite{braunstein,lloyd} where some few-qubits error-correcting codes were converted into few-modes CV error-correcting codes, no significant progress has been made in this direction. Although it was shown in Ref.~\cite{braunstein-nature} that these codes may be implemented with linear optics only, the type of errors that are correctable is arguably artificial (only one or a few modes may undergo noise, all others being required to suffer no errors at all, which is not realistic in an infinite-dimensional space).

In this Letter, we attack this problem from a different perspective, considering schemes to eliminate losses instead of noise 
in a CV quantum channel. We devise a CV quantum erasure-correcting code, which protects coherent states of light against probabilistic losses, or ``erasures'' in the qubit terminology. The protocol ensures the reliable transmission of coherent states over a channel that either transmits information perfectly or erases it completely with probability $p_e$. The channel thus transforms a coherent state $|\alpha\rangle$ into $\rho_{\alpha}=(1-p_e)|\alpha\rangle\langle\alpha|+p_e|0\rangle\langle0|$. Such a non-Gaussian loss model is known to occur in realistic situations, e.g., resulting from time jitter or beam pointing noise in atmospheric transmissions \cite{wittmann}. We will first show that if one can detect whether an erasure has occurred, our code allows one to correct it almost perfectly. Then, we will show how, using postselection, one can relax this requirement and still recover the original state with high fidelity. The resulting protocol nicely complements the techniques that have recently been developed to fight noise in CV quantum channels, including the purification of coherent states \cite{andersen} and squeezed states \cite{Heersink,franzen} from noisy copies, or the filtering of vacuum noise from an arbitrary set of coherent states \cite{wittmann}.

\begin{figure*}[t]
	\includegraphics[scale=0.37]{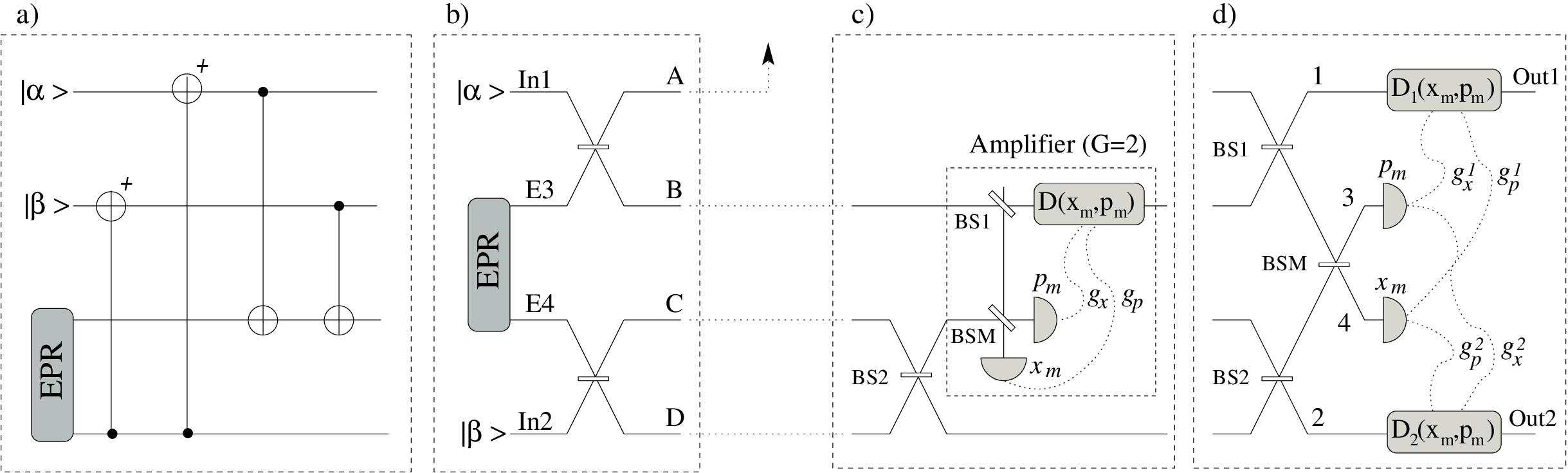} 
	\caption{a) Encoding circuit of the CV quantum erasure-correcting code; b) Optical implementation of this encoder using a two-mode vacuum squeezed state (EPR); c) Correction of an erasure of mode A via the phase-insensitive amplification of mode B realized with homodyne detection and feedforward; d) Decoding circuit correcting an erasure of any of the four modes.}
	\label{fig:fig1}
\end{figure*}

The erasure channel for qubits was first considered in Ref.~\cite{grassl}, where a quantum code protecting two qubits from erasure was devised, based on the encoding into a 4-qubit entangled state. The encoder quantum circuit is made of four control-NOT (CNOT) gates, and has been used, e.g., in the proposal for an all-optical quantum memory \cite{gingrich}. 
This circuit can be formally translated to CV by introducing the continuous-variable CNOT gate and its inverse (CNOT$^\dagger$) \cite{braunstein}, as shown in Fig.~\ref{fig:fig1}a. The resulting circuit can be turned into an optical scheme by using Bloch-Messiah reduction theorem, which states that a multimode evolution with linear Bogoliubov transformation 
$ \hat{b}_j = \sum_{k} (A_ {jk}\hat{a}_k + B_{jk} \hat{a}^{+}_k)$,
where $\hat{a}_j, \hat{b}_j$ are bosonic annihilation operators, may be decomposed into a multi-port linear interferometer followed by the parallel application of a set of single-mode squeezers, followed yet by another interferometer \cite{braunstein2}. 
After simplifications, we are left with the optical circuit of Fig.~\ref{fig:fig1}b, which boils down to mixing 
two input coherent states with an Einstein-Podolsky-Rosen (EPR) pair, in practice a two-mode squeezed vacuum state, at two balanced beam splitters. Note that a subpart of this circuit, where a coherent state is mixed with one beam of an EPR pair, has been introduced in the context of CV quantum secret sharing \cite{Tyc,Lance}. In this language, our encoder can be viewed as a (3,4) secret sharing protocol.

Let us prove now that one can correct losses provided that one monitors the occurrence of erasures. Depending on the channel, this monitoring may be achieved, e.g., by sending a probe pulse in an orthogonal mode, like another polarization, another spatial, or another frequency mode. Suppose we loose mode A during the transmission (see Fig. \ref{fig:fig1}b). We can recover the input coherent state $\vert\beta\rangle$ by mixing modes C and D on a balanced beam splitter, thus effectively completing a Mach-Zehnder interferometer. The other output port of the interferometer yields one half of the EPR pair. The recovery of the other state $\vert\alpha\rangle$ is a little more demanding as the information has been attenuated and polluted by quantum noise. However, this noise is exactly correlated with the  other half of the EPR pair, so that one can partly recover $\vert\alpha\rangle$ by amplifying mode B in a phase-insensitive amplifier of gain~2, using the second output port of the Mach-Zehnder interferometer as the idler input of the amplifier. Such an optical amplifier can be implemented using only linear optics, homodyne detection, and feedforward, as demonstrated in \cite{andersen2}. The decoder that corrects the loss of A based on this amplifier without nonlinearity is depicted in Fig.~\ref{fig:fig1}c.
Now, to have a practical protocol, the decoding should work regardless the location of the erasure. This is made possible by noticing first that the amplifier of Fig.~\ref{fig:fig1}c treats both input ports of BS1 on the same footing. Thus, if we connect A to the empty input of BS1 and adapt the sign of the electronic gains of the feedforward, the circuit can correct both erasures of A or B. Next, notice that BS1 now plays the same role for A and B as BS2 does for C and D. We thus find the optical circuit for the decoder shown in Fig.~\ref{fig:fig1}d.

Let us detail the protocol. For two input modes characterized by the conjugate quadrature operators $(\hat{x}_{in1},\hat{p}_{in1})$ and $(\hat{x}_{in2},\hat{p}_{in2})$, and an EPR pair corresponding to $(\hat{x}_{E3}-\hat{x}_{E4})/\sqrt{2}$ and $(\hat{p}_{E3}+\hat{p}_{E4})/\sqrt{2}$ being squeezed with variance $e^{-2r}$, the two output modes can be written as
\begin{align}\label{formula}
\nonumber \hat{x}_{out1(2)}&=\hat{x}_{1(2)}+g^x_{1(2)} \, \hat{x}_m \\
\hat{p}_{out1(2)}&=\hat{p}_{1(2)}+g^p_{1(2)} \, \hat{p}_m 
\end{align}
where $(\hat{x}_{1},\hat{p}_{1})$ and $(\hat{x}_{2},\hat{p}_{2})$ are the upper and lower output modes just before displacement, and $(\hat{x}_m,\hat{p}_m)$ are the measured quadratures. If we choose the electronic gains as indicated in Table~\ref{table}, one can easily check that the decoder yields one of the input coherent states with unit fidelity and the other one with a fidelity of
\begin{align}\label{fid1}
F=\frac{1}{1+e^{-2r}}. 
\end{align}
To verify it, suppose that mode A is lost during the transmission. The upper mode before displacement is given by
\begin{align}
\nonumber \hat{x}_{1}&=\frac{1}{\sqrt{2}} \hat{x}_v + \frac{1}{2} \hat{x}_{in1} - \frac{1}{2} \hat{x}_{E3}\\
\hat{p}_{1}&=\frac{1}{\sqrt{2}} \hat{p}_v + \frac{1}{2} \hat{p}_{in1} - \frac{1}{2} \hat{p}_{E3}
\end{align}
where $(\hat{x}_v,\hat{p}_v)$ refers to the vacuum mode introduced by the loss of A. The measured quadratures are given by
\begin{align}
\nonumber \hat{x}_m &=\frac{1}{2}\hat{x}_v - \frac{1}{2\sqrt{2}} \hat{x}_{in1} + \frac{1}{2\sqrt{2}} \hat{x}_{E3} - \frac{1}{\sqrt{2}} \hat{x}_{E4}\\
\hat{p}_m &=\frac{1}{2}\hat{p}_v - \frac{1}{2\sqrt{2}} \hat{p}_{in1} + \frac{1}{2\sqrt{2}} \hat{p}_{E3} + \frac{1}{\sqrt{2}} \hat{p}_{E4}
\end{align}
so that Eq.~\ref{formula} yields
\begin{align}
\nonumber \hat{x}_{out1}&=\hat{x}_{1}-\sqrt{2}\, \hat{x}_m = \hat{x}_{in1} - (\hat{x}_{E3}-\hat{x}_{E4})\\
\hat{p}_{out1}&=\hat{p}_{1}-\sqrt{2}\, \hat{p}_m = \hat{p}_{in1} - (\hat{p}_{E3}+\hat{p}_{E4})
\end{align}

\begin{table}[t]
\begin{tabular}{|c||cc|}
\hline
&($g^x_1$,$g^p_1$)&($g^x_2$,$g^p_2)$\\
\hline\hline
loss of A&$(-\sqrt{2}$,$-\sqrt{2})$&(0,0)\\
loss of B&($\sqrt{2}$,$\sqrt{2}$)&(0,0)\\
loss of C&(0,0)&($\sqrt{2}$,$-\sqrt{2}$)\\
loss of D&(0,0)&($-\sqrt{2}$,$\sqrt{2}$)\\
\hline
\end{tabular} 
\caption{Electronic gains for different loss locations.}
\label{table}
\end{table} 

One thus recovers the upper input with the fidelity of (\ref{fid1}), while the lower input is perfectly reconstructed via the lower Mach-Zehnder interferometer, that is, $(\hat{x}_{2},\hat{p}_{2})= (\hat{x}_{in2},\hat{p}_{in2})$.
Let us make a few comments here. First, these fidelities can be symmetrized by mixing the input modes entering the encoder and unmixing them at the output of the decoder, thus effectively distributing the added noise on both output modes. Next,
the fidelity (\ref{fid1}) -- or its symmetrized version -- is independent of the input coherent states, hence our scheme is universal. 
Finally, the decoder becomes perfect at the limit of infinite squeezing ($r\to\infty$).

Suppose now that we still send coherent states through a channel with probabilistic losses, but cannot longer probe erasures. In this more realistic situation, we do not know which set of gains to choose from Table~\ref{table} since the occurrence and location of erasures are unknown. In addition, we must consider multiple erasures, a possibility that was implicitly ignored above. As we shall see, our protocol can nevertheless be adapted to enable the transmission of coherent states immune to erasures provided that the deterministic feedforward is replaced by a probabilistic method based on postselection. The key idea is that if the measured quadratures are close to zero, then the output states do not need to be displaced regardless of the location of the erasure, i.e. all 4 lines of Table~\ref{table} imply the same action. Otherwise the output states must be discarded. This probabilistic protocol can thus be viewed as an erasure filter, which excludes the output states that have been affected by an erasure during transmission.

To investigate such a postselection, let us write the Wigner function of the two input modes carrying information together with the two modes of the EPR pair,
\begin{align}
W_{in}(r)=\frac{1}{\pi^4\sqrt{\mathrm{det} \gamma_{in}}}\exp[-(r-d_{in})\gamma_{in}^{-1}(r-d_{in})] 
\end{align}
where $r=(x_1,p_1,...,x_4,p_4)$ is the vector of quadrature components, $d_{in,j}=\langle r_j \rangle$ is the coherent vector, and $\gamma_{in,ij}=\langle r_i r_j+r_j r_i \rangle -2d_{in,i}d_{in,j}$ is the covariance matrix. This 4-mode state is processed through two parallel (lossy) Mach-Zehnder interferometers, then modes 3 and 4 are mixed on a balanced beam splitter and measured. Just before measurement, the 4-mode state will have evolved into a non-Gaussian mixture of Gaussian states, whose Wigner function can be written as $W_{out}(r)=\sum_{i=1}^{16} p_i W_{out}^{(i)}(r)$ with $W_{out}^{(i)}$ being the output Wigner function corresponding to one of the sixteen events that can occur during transmission. These events range from no erasure, with a probability of $(1-p_e)^4$, to the erasure of all four modes, with a probability of $p_e^4$. Next, the $p$ quadrature of mode 3 and $x$ quadrature of mode 4 are measured. If the outcomes are $(x_m,p_m)$, the Wigner function of the remaining modes reads
\begin{align}
\nonumber W_{out}(r'|x_m,p_m)&=\iint_{-\infty}^{-\infty} dx_3 dp_4 W_{out}(r',x_3,p_m,x_m,p_4)\\
&=\sum_{i=1}^{16} p_i \, W^{(i)}_{out}(r'|x_m,p_m)
\end{align}
where $r'=(x_1,p_1,x_2,p_2)$. To calculate these sixteen Wigner functions, we partition each covariance matrix $\gamma^{(i)}$ of the function $W_{out}^{(i)}$ before measurement with respect to the (traced over) quadratures $x_3$ and $p_4$.
We further partition the inverse of the covariance submatrix $\gamma'$ so that its block $\gamma''$ contains the second moments of the remaining modes after measurement, namely
\begin{equation}
\gamma^{(i)}=
\left(
\begin{array}{cc}
\gamma' & A\\
A^T & B
\end{array}
\right) \hspace{5mm}
(\gamma')^{-1}=
\left(
\begin{array}{cc}
(\gamma'')^{-1} & E\\
E^T & D
\end{array}
\right)
\end{equation}
After some calculations, we obtain 
\begin{align}
W^{(i)}_{out}(r'|x_m,p_m)&=\frac{1}{\pi^3 \sqrt{\mathrm{det} \gamma'}}\exp[-\delta^T F \delta]\\
\nonumber &\times \exp[-(r'-d')^T \gamma''^{-1}(r'-d')]
\end{align}
where $\delta$ is the vector of difference between the measured values $(x_m,p_m)$ in modes 3 and 4 and the corresponding mean values before measurement, $F=D-E^T \gamma''E$, and $d'=d_r-\gamma'' E \delta$, with $d_r$ being the coherent vector of modes 1 and 2 before displacement.

We now introduce a threshold condition, that is, we keep the output state only if $|x_m|\leq X_{th}$ and $|p_m|\leq P_{th}$. The resulting unnormalized Wigner function reads
\begin{align}\label{wigout}
W_{th}(r')=\sum_{i=1}^{16} p_i\int_{th} dx_m \, dp_m \, W^{(i)}_{out}(r'|x_m,p_m)
\end{align}
The probability to keep the output state is found by integrating (\ref{wigout}) over the phase space of the two output modes, i.e., $P_s=\int d^4r' \, W_{th}(r')$. To evaluate the quality of the protocol, we calculate the single-mode fidelity of one of the two output modes (say, mode~1), $F_{ps}=(2\pi/P_s) \int d^2r'_1 \, W^1_{th}(r'_1) \, W_0(r'_1)$, where $W_0$ is the Wigner function of the coherent state at input~1 and $W^1_{th}(r'_1)=\int d^2 r'_2 \, W_{th}(r')$. We then compare this fidelity to that resulting from the same  state being sent directly through the erasure channel. Note that $F_{ps}$ is state-dependent here, since the ability of the protocol to detect an erasure depends on the intensity of the two input beams. We nevertheless note that this dependence significantly affects the fidelity only at low intensities.

\begin{figure}
	\includegraphics[scale=0.265]{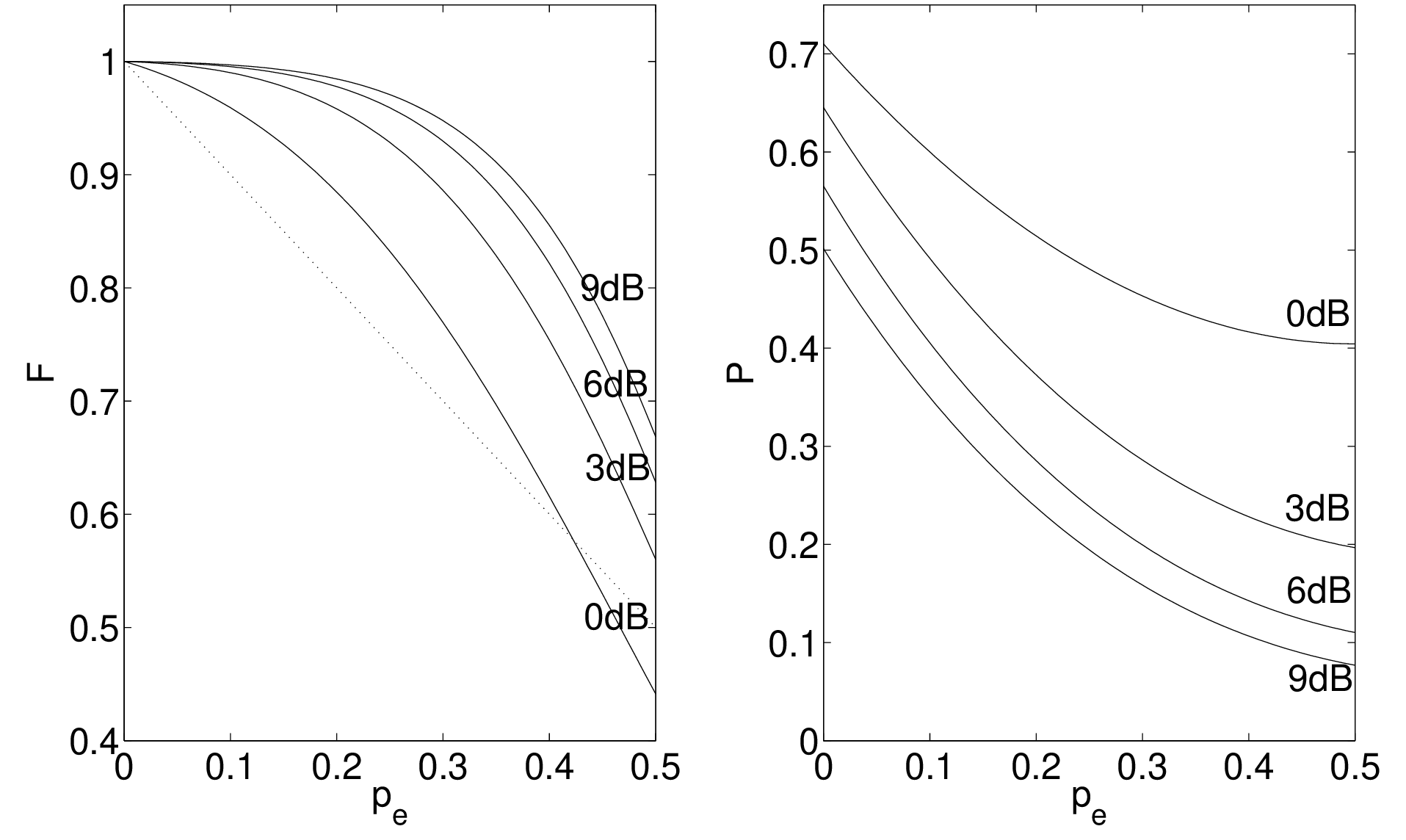}
	\caption{Single-mode fidelity (left) and success probability (right) versus the erasure probability for the input state $|\frac{4+i4}{\sqrt{2}}\rangle|0\rangle$. The solid lines correspond to various degrees of squeezing $e^{-2r}$, and $X_{th}=P_{th}=e^{-r}$. The dashed line is the fidelity when $|\frac{4+i4}{\sqrt{2}}\rangle$ is sent directly through the channel. All curves are plotted assuming $\eta_{HD}=0.9$ and $n_{e}=0$.}
	\label{fig:figure3}
\end{figure}

The performance of this erasure-filtering protocol is illustrated in Fig.~\ref{fig:figure3} for an input state $|(4+i4)/\sqrt{2}\rangle|0\rangle$ and various degrees of squeezing. Although the fidelity improves with squeezing, as expected, we observe that squeezing is not necessary. Interestingly, when no squeezing is used and one of the input states is vacuum, our scheme boils down to a very simple setup: the input coherent state is split on a balanced beam splitter, the two resulting modes are sent through the channel and interfere at the reception station. One of the two output beams is then heterodyne measured, and the other is kept conditionally on the outcomes being close to zero. The 0~dB curve of Fig.~\ref{fig:figure3} shows that this strikingly simple protocol is sufficient to allow an improved transmission of coherent state over the erasure channel.

Let us now foresee the experimental realization of the proposed protocol and address its feasibility. 
The efficiency of erasure filtering basically falls back on the quality of the entanglement source.
Gaussian entanglement can be produced through the interference of two Gaussian, single-mode squeezed states generated either using optical parametric oscillators \cite{entangled} or single-mode fibers \cite{glockl}. To enable high efficiency and self-locked interference between the two modes, we envisage a system where the two squeezed modes are produced in the same squeezing device but in orthogonal polarization modes. By using two orthogonally orientated nonlinear crystals inside a single cavity, the two polarization modes will be independently squeezed, have a relative phase which is inherently stable, and excite the same spatial mode as supported by the cavity \cite{lassen}. Using such a scheme, 6dB two-mode squeezing should be feasible. The outputs of the entanglement source must then interfere with two coherent states that can be defined as frequency sideband modes in a frequency range in which the entanglement is most pronounced. The resulting four beams are then mixed on three beam splitters. The spatial and temporal mode overlap at these beam splitters can be almost ideal by using a continuous-wave light source in a single spatial mode and a cavity based squeezing source. 

For the measurement of modes 3 and 4, one should use high efficiency and low noise homodyne detectors. To avoid the use of two separate local oscillators (one for each homodyne detector) a simpler scheme relying solely on two high sensitivity detectors can be employed, as discussed in \cite{andersen2}. The measurement efficiency $\eta_{HD}$ can then easily exceed 90\%. Furthermore, the electronic noise $n_{e}$ of the detectors and the associated feedforward electronics should be kept low. Electronic noise 2-3 orders of magnitude smaller than the shot noise is attainable \cite{andersen2}. 

In the deterministic scheme, the photocurrents must drive modulators traversed by auxiliary beams which subsequently are mixed with the output states 1 and 2 at very asymmetric beam splitters, thereby accomplishing a clean and near loss free displacement~\cite{teleportation,Tyc,andersen2}. In the probabilistic scheme, the analog outputs of the measurement devices should be digitized with a high-resolution analog-digital converter, providing fast measurements even when the success rate is low. The resulting outcome ($x_m,p_m$) is compared with the threshold values and the two output states are either selected or discarded. This selection process can be done electro-optically requiring fast real-time feedforward and fast amplitude modulators or, alternatively, pure electronically by selecting the digitized outcomes of the homodyne detectors used to characterize the scheme. 

To conclude, we stress that our protocol does not restrict to complete losses and coherent input states. Partial losses can be corrected as well, and the scheme applies, e.g., to the transmission of squeezed states over an erasure channel. Finally, the connection between error-correction and entanglement purification suggests an interesting extension of the proposed scheme to distribute entanglement over probabilistic lossy channels. We therefore expect our protocol to play an important role in the rapidly developing field of quantum communications.

We thank J. Fiurasek for useful discussions and acknowledge financial support from the EU under projects COVAQIAL and QAP.  
J.N. acknowledges support from the Belgian FRIA foundation.

\end{document}